\documentclass[10pt]{article}
\usepackage{fullpage}
\usepackage{amsmath}
\usepackage{amssymb}
\usepackage[dvips]{epsfig}
\usepackage{color}

\def\##1{{\bf #1}}
\def\=#1{\underline{\underline #1}}

\def\eps{\epsilon}

\def\ko{k_0}

\def\etao{\eta_0}
\def\.{\mbox{ \tiny{$^\bullet$} }}

\def\ux{\#{u}_x}
\def\uy{\#{u}_y}
\def\uz{\#{u}_z}

\def\un{\#{u}_n}
\def\ut{\#{u}_\tau}
\def\ub{\#{u}_b}

\def\le{\left(}
\def\ri{\right)}
\def\les{\left[}
\def\ris{\right]}
\def\lec{\left\{}
\def\ric{\right\}}

\def\c#1{\cite{#1}}
\def\l#1{\label{#1}}
\def\r#1{(\ref{#1})}

\begin{document}

\begin{center}

{\bf {\LARGE Modeling  columnar thin films as platforms for
surface--plasmonic--polaritonic optical sensing}}

\vspace{10mm} \large

 Tom G. Mackay\footnote{E--mail: T.Mackay@ed.ac.uk.}\\
{\em School of Mathematics and
   Maxwell Institute for Mathematical Sciences\\
University of Edinburgh, Edinburgh EH9 3JZ, UK}\\
and\\
 {\em NanoMM~---~Nanoengineered Metamaterials Group\\ Department of Engineering Science and Mechanics\\
Pennsylvania State University, University Park, PA 16802--6812,
USA}\\
 \vspace{3mm}
 Akhlesh  Lakhtakia\footnote{E--mail: akhlesh@psu.edu}\\
 {\em NanoMM~---~Nanoengineered Metamaterials Group\\ Department of Engineering Science and Mechanics\\
Pennsylvania State University, University Park, PA 16802--6812, USA}

\normalsize

\vspace{15mm}
{\bf Abstract}

\end{center}

\vspace{4mm} Via exploitation of surface plasmon polaritons (SPPs),
columnar thin films (CTFs) are attractive potential platforms for
optical sensing  as their relative permittivity dyadic and porosity
can be tailored to order. Nanoscale model parameters of a CTF  were
determined from its measured relative permittivity dyadic, after
inverting the Bruggeman homogenization formalism. These model
parameters were then used to determine the relative permittivity
dyadic of a fluid--infiltrated CTF. Two boundary--value problems
were next solved: the first  relating to SPP--wave propagation
guided by the planar interface of a semi--infinitely thick metal and
a semi--infinitely thick CTF, and the second to the plane--wave
response of the planar interface of a finitely thick metallic layer
and a CTF in a modified Kretschmann configuration. Numerical studies
revealed that   SPP waves propagate at a lower
  phase speed and with a shorter propagation length, if the fluid has a larger refractive index.
Furthermore, the angle of incidence required to excite an SPP wave
  in a modified Kretschmann configuration  increases as the  refractive index of the fluid increases.

\vspace{5mm}


\section{Introduction}

Research on optical sensors exploiting surface plasmon polaritons
(SPPs) has recorded an explosive growth
\cite{Green2000,Homola2003,Abdulhalim2008a} and commercial success
\cite{Arya2006,Luong2008} during the last two decades. SPPs are
virtual particles \c{Did} that can propagate guided by the planar interface
of a metal and a dielectric material, the classical analog of a
train of SPPs being an SPP wave.

Although several different sensing configurations have been
investigated to excite SPP waves, perhaps the most commonplace
 is the Kretschmann configuration \cite{Kret}. Both the metal and its dielectric partner in this configuration are layers of finite thickness. The metal film's thickness is $\sim100$~nm, whereas the dielectric layer has to be much thicker. On the other side of the metal film is a dielectric coupling material (in the form of a prism), which is optically denser than the dielectric material. Quasi-monochromatic light is launched at an angle to the thickness direction in the coupling material towards the metal film.  The fraction of illuminating light that is neither reflected nor transmitted is absorbed. As the angle of incidence increases from $0 $ towards $\pi/2$, a sharp peak in absorbance, accompanied by minuscule reflectance and transmittance, indicates the excitation of an SPP wave.  This sharp peak occurs only for $p$-polarized light,
and its angular location shifts when the refractive index of the dielectric partnering layer is altered by the incorporation
of an analyte.

The analyte can be incorporated in a variety of ways. One is to use a liquid such as water as the dielectric partner. Either the dispersal of the analyte in that liquid or the binding of the analyte to recognition molecules attached to the metal film can lead to an angular shift \cite{Homola2003,Abdulhalim2008a}. Another way is to use a highly
porous, thin metal film \cite{Yang1991,Maaroof2007}, so that it can be infiltrated by an analyte-containing liquid \cite{Abdulhalim2008b,Shalabaney}.

The dielectric partnering layer need not have an effectively isotropic constitution for the excitation of SPP waves. Uniaxial
dielectric materials have been extensively considered for the role of the dielectric partnering
material, both theoretically and experimentally
\cite{Sprokel1,Sprokel2,Lloyd1988,Kano,Baba}. Another class of candidates for this role is that of columnar thin films
(CTFs), which are effectively biaxial dielectric materials \cite{HW}.

A columnar thin film  is an assembly of parallel nanowires, grown on a planar substrate
by physical vapor deposition \c{HW,STF_Book}. As schematically
illustrated in Fig.~\ref{schematic1}, the column inclination angle $\chi \in \le 0, \pi/2 \ris$ is equal to or greater than
 the  vapor incidence angle $\chi_v \in \le 0, \pi/2 \ris$.
By control of the deposition-process parameters---the  vapor incidence angle in particular---and the material(s) evaporated,
both the optical properties and the porosity
of CTFs may be tailored to order, thereby controlling the
excitation conditions for SPP waves \cite{LP_AJP,PL_OC}. This renders CTFs attractive as potential platforms for
optical sensing of chemical and/or biological species that may infiltrate the void regions of the CTF \c{Barreca,PH07}.

Our aim for the study reported here was to determine the sensitivity
of SPP--wave excitation at the planar interface of a metal and a CTF
to changes in the refractive index of the fluid infiltrating the
void regions of the CTF. Employing the relative permittivity dyadic
of an uninfiltrated CTF, this study is largely numerical. It is
primarily based  on theoretical results established in two recent
papers: one on the theory of exciting an SPP wave that propagates
parallel to the morphologically significant plane of the CTF
\c{LP_AJP}, the second on inverting the Bruggeman homogenization
formalism with data on the uninfiltrated CTF \c{ML_inverse_homog}
for use in a forward Bruggeman formalism \c{Lakh_Opt} to compute the
effective relative permittivity dyadic of the infiltrated CTF.

The plan of this paper is as follows: Section~\ref{homo} succinctly describes the inverse and forward
Bruggeman homogenization formalisms for the uninfiltrated and the infiltrated CTF. Two boundary--value
problems are formulated and solved next:
\begin{itemize}
\item[(i)] The first is a canonical problem relating to SPP--wave propagation guided by the planar interface of a
semi--infinitely thick metal and a semi--infinitely thick CTF (Sec.~\ref{SWP}), and
\item[(ii)] the second is a realistic problem
involving the plane--wave response of the planar interface of a finitely thick metallic layer
and a CTF in a modified Kretschmann configuration (Sec.~\ref{MKC}).
\end{itemize}
Detailed descriptions of
theoretical formulations are available in the predecessor papers \c{LP_AJP,ML_inverse_homog}.
In the notation adopted, 3$\times$3 dyadics are double underlined while 3--vectors are in boldface;
unit vectors along the Cartesian axes are denoted $\ux$, $\uy$ and $\uz$.
 The operators $\mbox{Re} \les \cdot \ris$ and
$\mbox{Im} \les \cdot \ris$ deliver the real and imaginary parts of complex--valued quantities; and $i = \sqrt{-1}$. The free--space wavenumber is $\ko$
and the  intrinsic impedance of free space is $\etao$. An $\exp(-i\omega t)$ time--dependence is implicit,
with $\omega$ as the angular frequency and $t$ as time.

\section{Homogenization studies on CTFs}\label{homo}

\subsection{Preliminaries}

Let us begin with an uninfiltrated CTF characterized at length scales
much greater than the nanoscale by the relative permittivity dyadic
\begin{equation}
\=\eps_{\,ctf1} = \eps_{a1}  \,\un\un +\eps_{b1}\,\ut\ut \,
+\,\eps_{c1}\,\ub\ub.
\end{equation}
Herein the normal, tangential and binormal basis vectors are
specified in terms of the column inclination angle $\chi \in \le 0,
\pi/2 \ris$ per
\begin{equation}
\left. \begin{array}{l}
 \un = - \ux \, \sin \chi + \uz \, \cos \chi \vspace{4pt} \\
 \ut =  \ux \, \cos \chi + \uz \, \sin \chi \vspace{4pt} \\
\ub = - \uy
\end{array}
\right\}.
\end{equation}
The  CTF is grown on a planar substrate, which we take to lie
parallel to $z=0$, through the deposition of an evaporated bulk
material. The vapor incidence angle $\chi_v$ is less than or equal
to the column inclination angle $\chi$, as schematically illustrated
in Fig.~\ref{schematic1}. The deposited material is assumed to be an
isotropic dielectric material with refractive index $n_s$. The morphologically
significant plane of the CTF is the $xz$ plane \cite[Chap. 7]{STF_Book}.

For the purposes of generating the numerical results presented in
the following sections,  we selected   CTFs made from evaporating
patinal${}^{\mbox{\textregistered}}$  titanium oxide. The relative
permittivity parameters of these CTFs are
\begin{equation}
\left.
\begin{array}{l}
\eps_{a1} = \displaystyle{\les 1.0443 + 2.7394 \le \frac{2
\chi_v}{\pi} \ri - 1.3697
\le \frac{2 \chi_v}{\pi} \ri^2 \ris^2} \vspace{6pt} \\
\eps_{b1} = \displaystyle{ \les 1.6765 + 1.5649 \le \frac{2
\chi_v}{\pi} \ri - 0.7825 \le \frac{2 \chi_v}{\pi} \ri^2 \ris^2}
 \vspace{6pt} \\
\eps_{c1} = \displaystyle{ \les 1.3586 + 2.1109 \le \frac{2
\chi_v}{\pi} \ri - 1.0554 \le \frac{2 \chi_v}{\pi} \ri^2 \ris^2}
\end{array}
\right\} \l{tio1}
\end{equation}
with
\begin{equation}
\tan \chi = 2.8818 \, \tan \chi_v, \l{tio2}
\end{equation}
as determined by experimental measurements at a free--space
wavelength of 633 nm  \c{HWH_AO}.

Each column of a CTF  can be modeled as an assembly of
elongated ellipsoidal inclusions, strung together end-to-end \cite[Chap. 6]{STF_Book}. The
inclusions have identical orientations and shapes. The inclusion surface is characterized  by
the  dyadic
\begin{equation}
 \un \, \un + \gamma_\tau \, \ut \, \ut + \gamma_b \, \ub \,
\ub.
\end{equation}
The highly aciculate nature of the columnar  morphology suggests
that the shape parameters $\gamma_{b} \gtrsim 1$ and $\gamma_\tau
\gg 1$. In fact, we fix $\gamma_\tau = 15$  since increasing
$\gamma_\tau$ beyond 10 does not bring about significant effects for
slender inclusions \cite{Lakh_Opt}.
The CTF's columns occupy only a proportion $f \in \le 0, 1 \ri $ of
the total CTF volume. The volume fraction of the CTF  not occupied
by columns is $1 - f$.

\subsection{Inverse Bruggeman formalism~---~uninfiltrated CTF} \l{InverseHomog}

While the relative--permittivity parameters $\lec \eps_{a1}, \eps_{b1}, \eps_{c1} \ric$
may be straightforwardly measured, the same is not true for the nanoscale model parameters
$\lec n_s, f, \gamma_b \ric$. The refractive index of the bulk material is known prior to evaporation,
but the refractive index of the deposited material may be quite different,
 depending on the precise details of the deposition environment \cite{MTR1976,BMYVM,WRL03}.
Also, the accurate determinations of $\gamma_b$ from scanning--electron micrographs \cite{HW}, and $f$ by gas--adsorption
techniques \cite{BET1938,BHBP,RHLP}
or
mass density measurements  \cite{MTR1976}, are known to be problematic.
Therefore, our first goal in this section is to establish the CTF nanoscale parameters
$\lec n_s, f, \gamma_b \ric$, from a knowledge of the eigenvalues
$\lec \eps_{a1}, \eps_{b1}, \eps_{c1} \ric$ of $\=\eps_{\,ctf1}$. The void regions
of the uninfiltrated CTF are taken to be vacuous.

The computation of $\lec n_s, f, \gamma_b \ric$ can be undertaken
using the process of  inverse homogenization. The widely used
Bruggeman homogenization formalism \c{WLM,EAB} provides an suitable framework
for this process. Since we have recently described this inverse
homogenization process in detail elsewhere \c{ML_inverse_homog}, here we need only
present the results.

For the titanium--oxide CTF described by the
constitutive parameters \r{tio1} and \r{tio2}, the corresponding
values of $\lec n_s, f, \gamma_b \ric$ are presented in
Table~\ref{tab1} for $\chi_v \in \lec 15^\circ, \, 30^\circ, \,
60^\circ, \, 90^\circ \ric$. The calculated data are in accord
with previous observations:
\begin{itemize}
\item As $\chi_v$ increases towards $\pi/2$, the biaxiality of
a CTF reduces towards uniaxiality \c{HW}, which is reflected by the tendency
of $\gamma_b$ in Table~\ref{tab1} to reduce to almost unity.

\item CTFs deposited more obliquely tend to have lower mass density \cite[Chap. 5]{STF_Book},\cite{Tait}.
This is confirmed Table~\ref{tab1} by the monotonic growth in the volume fraction $f$ as
$\chi_v$ increases towards $\pi/2$.

\end{itemize}

We also noticed a steady decrease in $n_s$ towards the bulk refractive index of
titanium oxide, as $\chi_v\to\pi/2$. This trend does depend on the
material deposited, and can even be turned into a steady increase
\c{ML_inverse_homog}.
It suggests the complexity of the roles played
by the nanoscale parameters $\lec n_s, f, \gamma_b \ric$ in the
emergence of the relative--permittivity parameters $\lec \eps_{a1},
\eps_{b1}, \eps_{c1} \ric$.

\subsection{Forward homogenization~---~infiltrated CTF}
\l{ForwardHomog}

In an optical sensor exploiting the SPP--wave phenomenon,  the void regions of the CTF
have to be filled with a
fluid of refractive index $n_\ell$. As a result,
$\=\eps_{\,ctf1}$   comprising the eigenvalues $\lec \eps_{a1}, \eps_{b1}, \eps_{c1}
\ric$ changes to $\=\eps_{\,ctf2}$
comprising new eigenvalues
$\lec \eps_{a2}, \eps_{b2}, \eps_{c2} \ric$; the column inclination angle $\chi$ remains
unchanged. From
knowledge of the nanoscale model parameters $\lec n_s,  f,
\gamma_b \ric$ as well as of $\lec n_\ell,\gamma_\tau\ric$, the Bruggeman homogenization formalism
can be applied in its usual forward sense to determine
$\lec \eps_{a2}, \eps_{b2}, \eps_{c2} \ric$.
A description of the forward Bruggeman formalism as applied to a CTF is
available elsewhere \c{Lakh_Opt}.

After using the nanoscale model parameters presented in Table~\ref{tab1},
the corresponding values of $\lec \eps_{a2}, \eps_{b2}, \eps_{c2}
\ric$ were  plotted against $n_\ell \in \le 1.0, 1.5 \ri$ in
Fig.~\ref{Fig1} for $\chi_v \in \lec 15^\circ, \, 30^\circ, \,
60^\circ, \, 90^\circ \ric$. We see that each of $\eps_{a2}$,
$\eps_{b2}$ and $\eps_{c2}$ increases approximately linearly as
$n_\ell$ increases. Also, the rate of increase is greater for
smaller values of $\chi_v$, because the CTF is then more porous.

\section{Canonical boundary--value problem} \l{SWP}

Let us now investigate the wavenumbers of SPP
waves guided by the planar interface of two half--spaces, one filled
with a metal and the other occupied by a CTF.
Since the underlying theory for this canonical problem is
described elsewhere \c{LP_AJP}, here we present only the key
theoretical results as a precursor to our numerical results.

Suppose that the half--space $z>0$ is occupied by a
fluid--infiltrated  CTF while a metal of relative permittivity
$\eps_m$ occupies the half--space $z<0$. The SPP wave under
consideration is $p$--polarized and propagates in the $xz$ plane. In
the metal half--space, the electromagnetic phasors are
\begin{equation}
\left. \begin{array}{l} \#E (\#r) = \displaystyle{A_m \le \ux -
\frac{i \sigma}{q_m} \uz \ri \, \exp \les i \ko \le \sigma x - i q_m
z
\ri \ris} \vspace{6pt} \\
 \#H (\#r) = \displaystyle{A_m \frac{i \eps_m }{\etao q_m} \uy \, \exp \les i \ko \le \sigma x - i q_m z
\ri \ris}
\end{array} \right\}, \qquad z \leq 0,
\end{equation}
where $A_m$ is the complex--valued amplitude, $q_m = \sqrt{\sigma^2
- \eps_m}$ and $\sigma \ko \ux$ represents the wave vector of the
SPP wave. By choosing $\mbox{Re} \,\les \, q_m  \, \ris > 0$, we
ensure that this  wave decays away from the interface $z=0$. The
corresponding electromagnetic field phasors in the
fluid--infiltrated--CTF half--space are
\begin{equation}
\left. \begin{array}{l} \#E (\#r) = \displaystyle{A_c \les \ux +
\frac{i \sigma q_c - \le \eps_{a2} - \eps_{b2} \ri \, \sin \chi \,
\cos \chi}{\sigma^2 - \le \eps_{a2} \cos^2 \chi + \eps_{b2} \sin^2
\chi \ri}\, \uz \ris \, \exp \les i \ko \le \sigma x + i q_c z
\ri \ris} \vspace{6pt} \\
 \#H (\#r) = \displaystyle{A_c \frac{  \sigma \le \eps_{a2} - \eps_{b2} \ri \,
  \sin \chi \, \cos \chi - i q_c \le \eps_{a2} \cos^2 \chi + \eps_{b2} \sin^2 \chi \ri }
  {\etao \les  \sigma^2 - \le \eps_{a2} \cos^2 \chi + \eps_{b2} \sin^2 \chi \ri \ris} \, \uy \, \exp \les i \ko \le \sigma x + i q_c z
\ri \ris}
\end{array} \right\}, \qquad z \geq 0,
\end{equation}
where $A_c$ is the complex--valued amplitude and the quadratic
equation  \c{LP_AJP}
\begin{equation}
q_c^2 \le \eps_{a2} \cos^2 \chi + \eps_{b2} \sin^2 \chi \ri + 2 i
\sigma q_c \le \eps_{a2} - \eps_{b2} \ri \, \sin \chi \, \cos \chi -
\sigma^2 \le \eps_{a2} \sin^2 \chi + \eps_{b2} \cos^2 \chi \ri +
\eps_{a2} \eps_{b2} = 0
\end{equation}
 yields $q_c$.
 The choice $\mbox{Re} \, \les \, q_c \, \ris  > 0$ ensures that
this  wave decays away from the interface.

Upon  applying the standard boundary conditions at $z=0$, the
dispersion relation  \c{LP_AJP}
\begin{equation}
\eps_m \sigma^2 + i q_m \sigma \le \eps_{a2} - \eps_{b2} \ri \, \sin
\chi \, \cos \chi + \le q_m q_c - \eps_m \ri \le \eps_{a2} \cos^2
\chi + \eps_{b2} \sin^2 \chi \ri =  0
\end{equation}
emerges, from which the  relative wavenumber $\sigma$ may be
extracted by numerical means.

For illustrative purposes, let us choose the metal occupying $z<0$
to be bulk aluminum, for  which $\eps_m = -56 + 21i$ at a
free--space wavelength of $633$ nm.
 In Fig.~\ref{Fig2}, the real and
imaginary parts of $\sigma$ are plotted against $n_\ell$ for $\chi_v
\in \lec 15^\circ, \, 30^\circ, \, 60^\circ, \, 90^\circ \ric$.
 The real part of $\sigma$ increases
approximately linearly as $n_\ell$ increases, and it also increases
as $\chi_v$ increases. From this we infer that the phase speed of
the SPP wave decreases as $n_\ell$ increases and as $\chi_v$
increases. Similarly,  the imaginary part of $\sigma$ also increases
as both $n_\ell$ and $\chi_v$ increase, and from this we infer that
the attenuation of  the SPP wave increases as $n_\ell$ increases and
as $\chi_v$ increases.

\section{Modified Kretschmann configuration}\label{MKC}

We now turn to a realistic setup for launching SPP waves along the
planar interface of a metal film and a CTF of finite thickness. We
examine a modification \cite{LP_AJP} of the Kretschmann
configuration \c{Kret} wherein the regions $z \leq 0$ and $z \geq
L_\Sigma$ are occupied by homogeneous, isotropic, nondissipative,
dielectric materials with relative permittivity scalars $\eps_{d}$
and $\eps_{\ell}=n_\ell^2$, respectively. The fluid--infiltrated CTF
 occupies the
laminar region $L_m \leq z \leq L_\Sigma$, while the laminar region
$0 \leq z \leq L_m$ is occupied by a metal with relative
permittivity $\eps_m$. A schematic representation of this modified
Kretschmann configuration is provided in Fig.~\ref{schematic2}.

Let us consider a $p$--polarized plane wave in the region $z \leq
0$, propagating towards the metal--coated CTF at an angle
$\theta_{inc} \in \le 0, \pi/2 \ri$ to the $z$ axis, as described by
the electromagnetic field phasors
\begin{equation}
\left.
\begin{array}{l}
\#E_{inc} = \le - \ux \cos \theta_{inc} + \uz \sin \theta_{inc} \ri
\, \exp \les i \le \kappa x + \sqrt{\eps_{d}}\, z \cos \theta_{inc}
\ri \ris \vspace{6pt}\\
\#H_{inc} = \displaystyle{ - \frac{\sqrt{\eps_{d}}}{\etao} \, \uy\,
\exp \les i \le \kappa x + \sqrt{\eps_{d}}\, z \cos \theta_{inc}
\ri \ris}
\end{array}
\right\}, \qquad z \leq 0,
\end{equation}
with
\begin{equation}
\kappa = \ko \sqrt{\eps_{d}} \, \sin \theta_{inc}.
\end{equation}
The corresponding reflected and transmitted electromagnetic field
phasors are given as
\begin{equation}
\left.
\begin{array}{l}
\#E_{ref} = r_p  \le  \ux \cos \theta_{inc} + \uz \sin \theta_{inc}
\ri \, \exp \les i \le \kappa x - \sqrt{\eps_{d}}\, z \cos
\theta_{inc}
\ri \ris \vspace{6pt}\\
\#H_{ref} = \displaystyle{ - \frac{r_p \sqrt{ \eps_{d}}}{\etao} \,
\uy\, \exp \les i \le \kappa x - \sqrt{\eps_{d}}\, z \cos
\theta_{inc} \ri \ris}
\end{array}
\right\}, \qquad z \leq 0,
\end{equation}
and
\begin{equation}
\left.
\begin{array}{l}
\#E_{tr} = t_p  \le - \ux \cos \theta_{tr} + \uz \sin \theta_{tr}
\ri \, \exp \les i \le \kappa x + \sqrt{\eps_{\ell}}\,\le z - L_\Sigma
\ri \, \cos \theta_{tr}
\ri \ris \vspace{6pt}\\
\#H_{tr} = \displaystyle{ - \frac{t_p \sqrt{ \eps_{\ell}}}{\etao} \,
\uy\, \exp \les i \le \kappa x + \sqrt{\eps_{\ell}}\, \le z - L_\Sigma
\ri \,\cos \theta_{tr} \ri \ris}
\end{array}
\right\}, \qquad z \geq L_\Sigma,
\end{equation}
respectively, where
\begin{equation}
\sqrt{\eps_{d}} \, \sin \theta_{inc} = \sqrt{\eps_{\ell}} \, \sin
\theta_{tr} = \sigma.
\end{equation}
The complex--valued reflection and transmission coefficients, namely $r_p$ and
 $t_p$, are determined by solving the related
boundary--value problem. For full details, the reader is referred
elsewhere \c{STF_Book,LP_AJP}.

In  studying surface waves at the metal--CTF interface, a key
parameter is the absorbance
\begin{equation}
A_p = 1 - \le  |\, r_p \, |^2 + \frac{\sqrt{\eps_{\ell}} \; \mbox{Re}
\, \les \, \cos \theta_{tr} \, \ris}{\sqrt{\eps_{d}} \, \cos
\theta_{inc}}\, | \, t_p \, |^2 \ri.
\end{equation}
A characteristic signature of SPP--wave excitation is a peak in graph of $A_p$
versus $\theta_{inc}$, when the angle of incidence exceeds the critical angle
in the absence of the metal film.

For our numerical study, we chose $L_m = 10$ nm and $L_\Sigma = L_m
+ 1000$ nm. The existence of  a critical angle for total reflection
in the absence of the metal film follows by setting the relative
permittivity $\eps_{d} = 2.6^2=6.76$, in the neighborhood of that
which can be delivered by a rutile prism. In contrast,  the relative
permittivity $\eps_{\ell}$ was kept variable. As in Sec.~\ref{SWP},
aluminum was chosen as the metal.

In Fig.~\ref{Fig3}, graphs of $A_p$ versus $\theta_{inc}$ are
presented for $n_\ell \in \lec 1.0, 1.25, 1.5 \ric$ with $\chi_v =
30^\circ$. Also plotted versus $\theta_{inc}$ is the quantity
$|\,r_p\,|^2$, computed when $L_m = 0$;  these plots exhibit an
abrupt step from $0$ to $1$, which occurs at the critical value of
$\theta_{inc}$ for total reflection.
 The rightmost peak in
the graphs $A_p$,
which occurs beyond the critical angle,
 indicates the excitation of SPP waves.

 The fairly sharp spikes in the graphs of $A_p$,
to the left of the rightmost peak, indicate bulk guided modes
\c{Welford1987}. The $\theta_{inc}$ values which correspond to these
modes depend upon the thickness of the CTF. To demonstrate this, the
plots of Fig.~\ref{Fig3}  are reproduced in Fig.~\ref{Fig3_L1500}
but with $L_{\Sigma} = L_m + 1500$ nm. From a comparison of
Figs.~\ref{Fig3} and \ref{Fig3_L1500}, it is clear that  bulk guided
modes  for $L_{\Sigma} = L_m + 1500$ nm arise at values of
$\theta_{inc}$ which are different to those for $L_{\Sigma} = L_m +
1000$ nm. Importantly, the values of $\theta_{inc}$ which correspond
to the excitation of SPPs are the same for $L_{\Sigma} = L_m + 1500$
nm and $L_{\Sigma} = L_m + 1000$ nm.

The value of $\theta_{inc}$ which corresponds to the rightmost peak in the graph of  $A_p$~---~let  us call
this value $\theta^\sharp_{inc} (n_\ell)$ at a specified value of
$n_\ell$~---~is clearly  sensitive to both $n_\ell$ and $\chi_v$. The  value $\theta^\sharp_{inc} (n_\ell)$
increases as $n_\ell$ increases and as $\chi_v$ increases.
In order to quantify the sensitivity of  $\theta^\sharp_{inc} (n_\ell)$, we introduce the
 figure of merit
\begin{equation}
\rho = \frac{\theta^\sharp_{inc} (n_\ell) - \theta^\sharp_{inc} (1.0)}{n_\ell
- 1.0},
\end{equation}
for $n_\ell \in (1.0, 1.5)$.
Graphs of $\rho$ versus $n_\ell$ are provided in Fig.~\ref{Fig4} for $\chi_v \in
 \lec 15^\circ, 30^\circ, 60^\circ, 90^\circ \ric$ and
$L_{\Sigma} = L_m + 1000$ nm. These clearly indicate that the
metal--coated CTF can function as an optical
 sensor for an analyte  dispersed uniformly in a solution. Furthermore,
we observe that $\theta^\sharp_{inc} (n_\ell)$ is most sensitive to changes in
 $n_\ell$ when $\chi_v$ is small and $n_\ell$ is large.

\section{Closing remarks}

Our numerical study has demonstrated that the excitation of an SPP wave
guided by a planar interface of a metal films and a CTF   is highly sensitive to the refractive index
 of a fluid infiltrating the CTF, as well as the morphology of the CTF itself.
 In particular, as the refractive index of the fluid increases, the phase speed of the SPP wave decreases and its degree of
 attenuation increases. Furthermore, the angle of incidence required to excite the
 SPP wave
  in a modified Kretschmann configuration  increases as the refractive index of the fluid increases.
This sensitivity bodes well for the implementation of
fluid--infiltrated CTFs, as well as fluid--infiltrated sculptured
thin films \cite{STF_Book} more generally,
 as SPP--based optical sensors.

\vspace{10mm}

\noindent {\bf Acknowledgments:} TGM is supported by a  Royal
Academy of Engineering/Leverhulme Trust Senior Research Fellowship.
AL thanks the Binder Endowment at Penn State for partial financial
support of his research activities.

\normalsize
 \newpage
\begin{table}[ht]
\begin{center}
\begin{tabular}{|c  | c | c | c|} \hline \vspace{-4pt} &&&   \\
$\chi_v$ & $\gamma_b$ & $f$ & $n_s$   \\ &&& \vspace{-4pt} \\  \hline &&& \vspace{-4pt} \\
$15^\circ$ & 2.2793 & 0.3614 & 3.2510 \\ \hline &&& \\
$30^\circ$ & 1.8381 & 0.5039 & 3.0517 \\ \hline &&& \\
$60^\circ$ & 1.4054 & 0.6956 & 2.9105 \\ \hline &&& \\
$90^\circ$ & 1.0020 & 0.7859 & 2.8228 \\ \hline
\end{tabular}
\caption{The dimensionless quantities $\gamma_b$, $f$ and $n_s$ computed  using
the inverse Bruggeman homogenization formalism for a titanium--oxide
CTF with $\chi_v = 15^\circ$, $30^\circ$, $60^\circ$ and
$90^\circ$.} \label{tab1}
\end{center}
\end{table}

\newpage

\begin{figure}[!ht]
\centering
\includegraphics[width=3.5in]{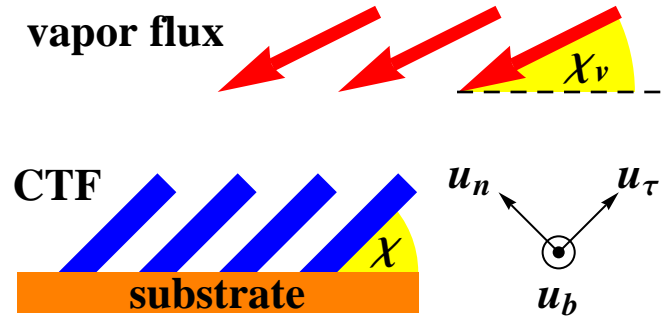}
 \caption{ \l{schematic1}
A columnar thin film  growing with column inclination angle $\chi$,
with the vapor incidence angle  $\chi_v \leq \chi$. The columns grow
along the direction of the unit vector $\#u_\tau$. }
\end{figure}

\newpage

\begin{figure}[!ht]
\centering
\includegraphics[width=3.5in]{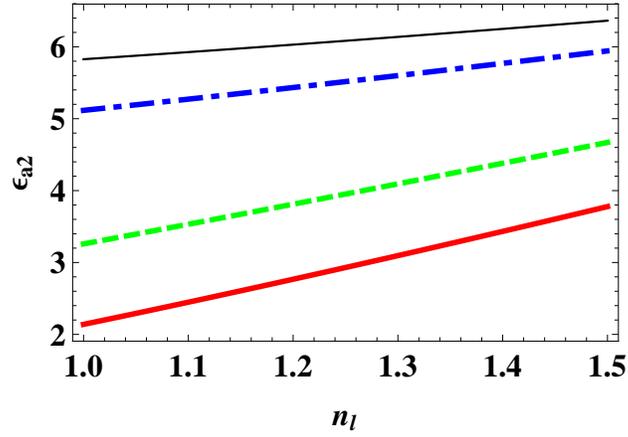}
\includegraphics[width=3.5in]{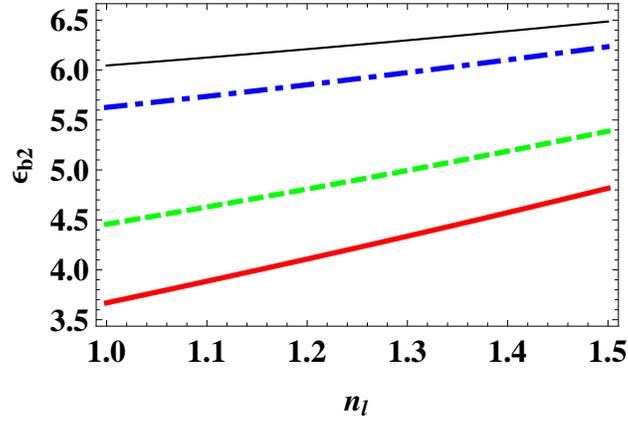}
\includegraphics[width=3.5in]{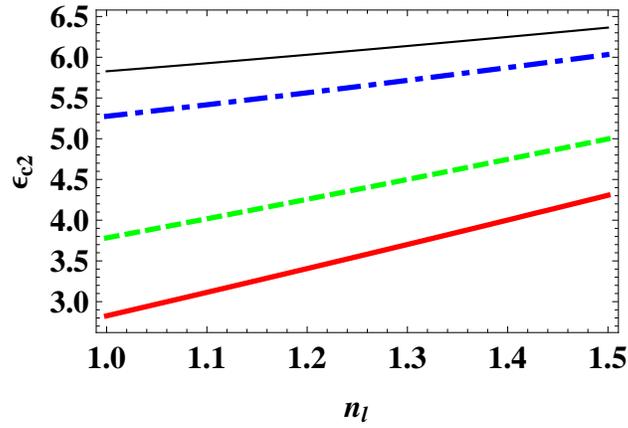}
 \caption{ \l{Fig1}
 The relative permittivity--parameters $\eps_{a2}$,  $\eps_{b2}$ and
 $\eps_{c2}$ of the fluid--infiltrated CTF, as computed using the forward Bruggeman homgenization formalism,
  plotted against the refractive index $n_\ell$ of the fluid infiltrating the void regions of the CTF,
for $\chi_v = 15^\circ$ (red, thick solid curve), $30^\circ$ (green, dashed curve), $60^\circ$ (blue, broken dashed
 curve) and $ 90^\circ$ (black, thin solid curve).}
\end{figure}

\newpage

\begin{figure}[!ht]
\centering
\includegraphics[width=3.5in]{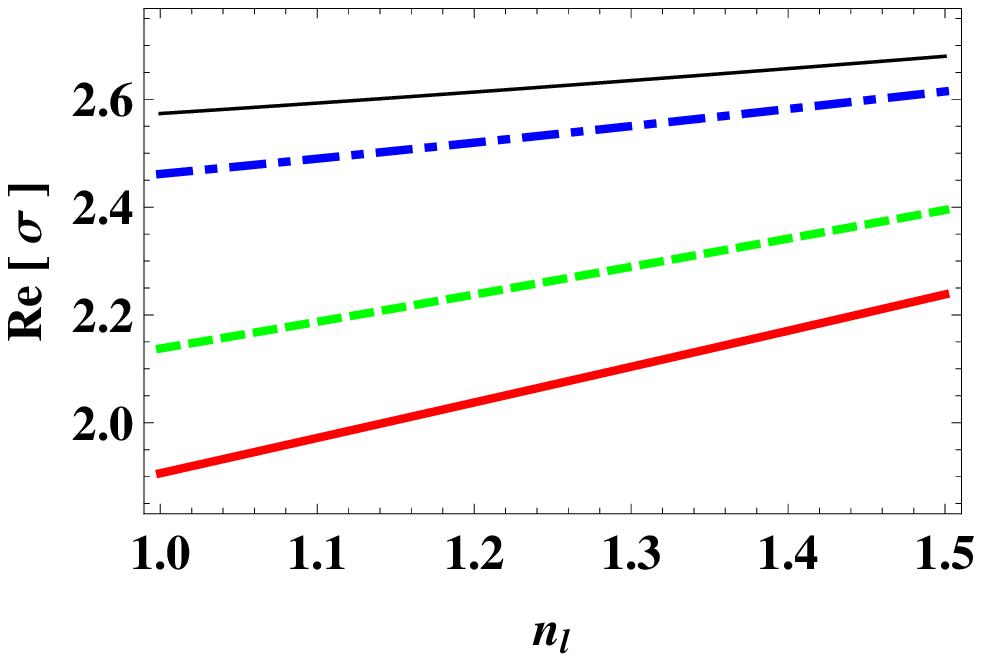}
\includegraphics[width=3.5in]{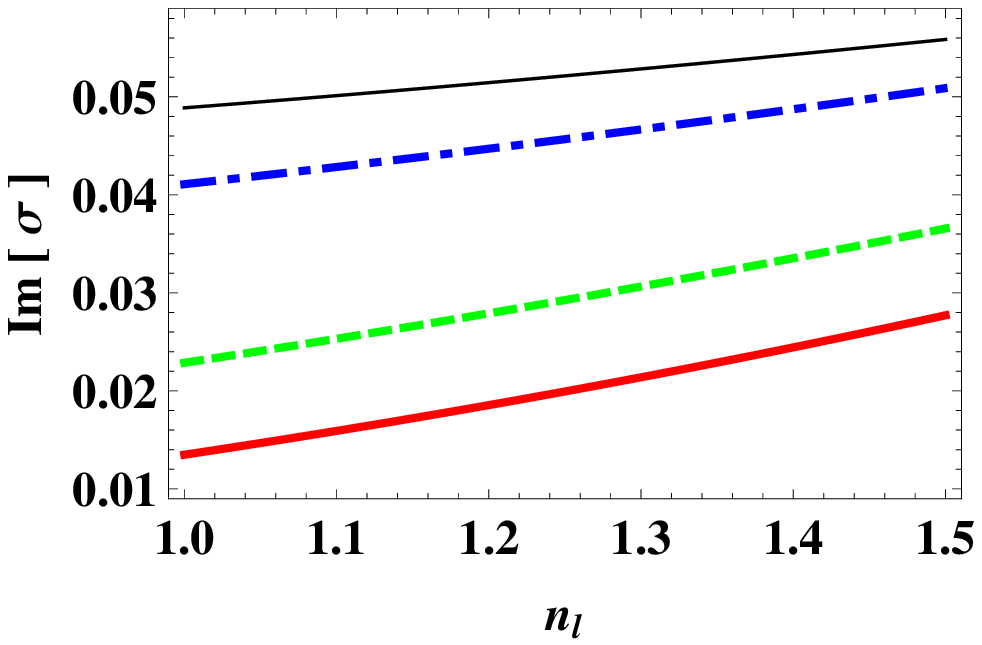}
 \caption{ \l{Fig2}
 The real and imaginary parts of $\sigma$
 plotted against  the refractive index $n_\ell$ of the fluid infiltrating the void regions of the CTF
  for $\chi_v = 15^\circ$ (red, thick solid curve), $30^\circ$ (green, dashed curve), $60^\circ$ (blue, broken dashed
 curve) and $ 90^\circ$ (black, thin solid curve). }
\end{figure}

\newpage

\begin{figure}[!ht]
\centering
\includegraphics[width=3.5in]{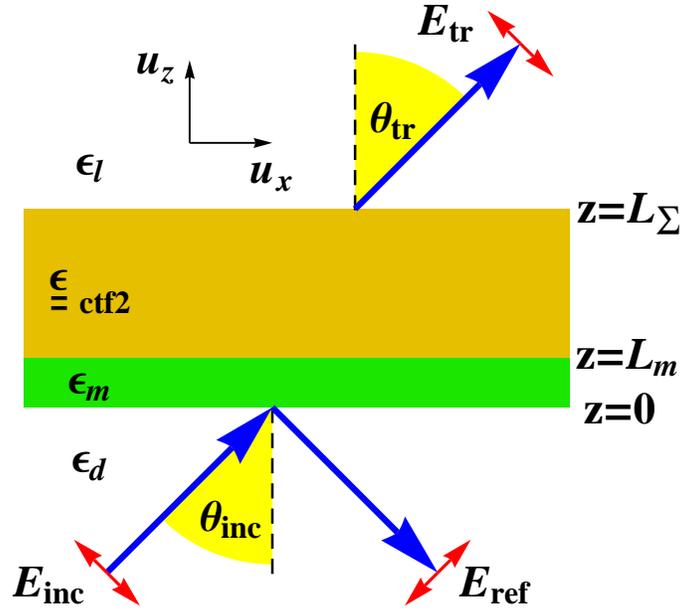}
 \caption{ \l{schematic2}
A $p$--polarized plane wave incident on a metal--coated CTF in the
modified Kretschmann configuration.}
\end{figure}

\newpage

\begin{figure}[!ht]
\centering
\includegraphics[width=3.05in]{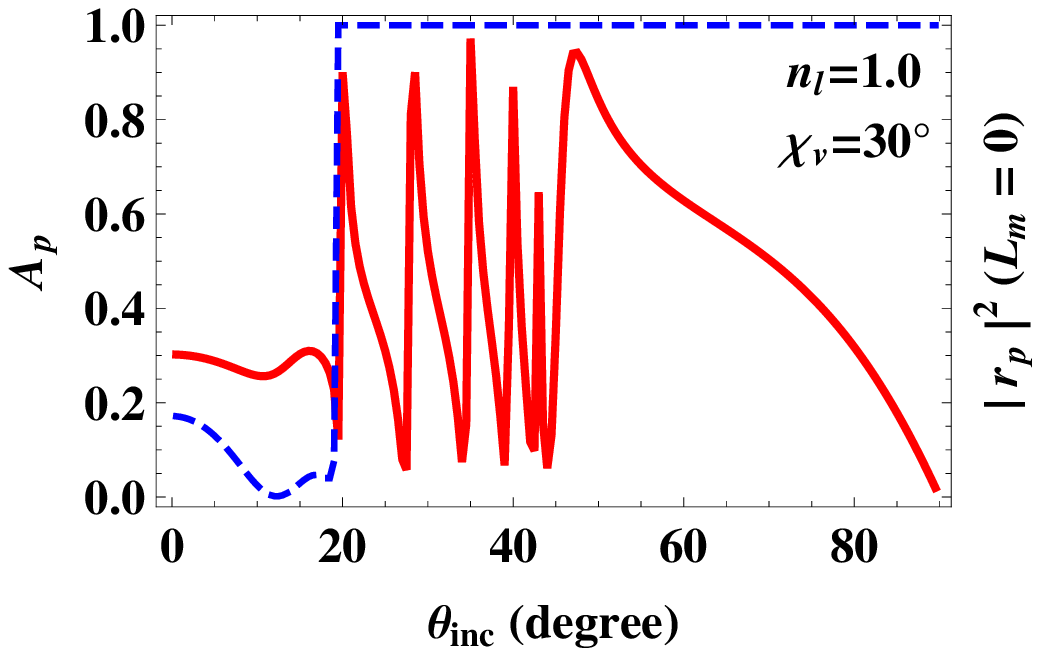}
\includegraphics[width=3.05in]{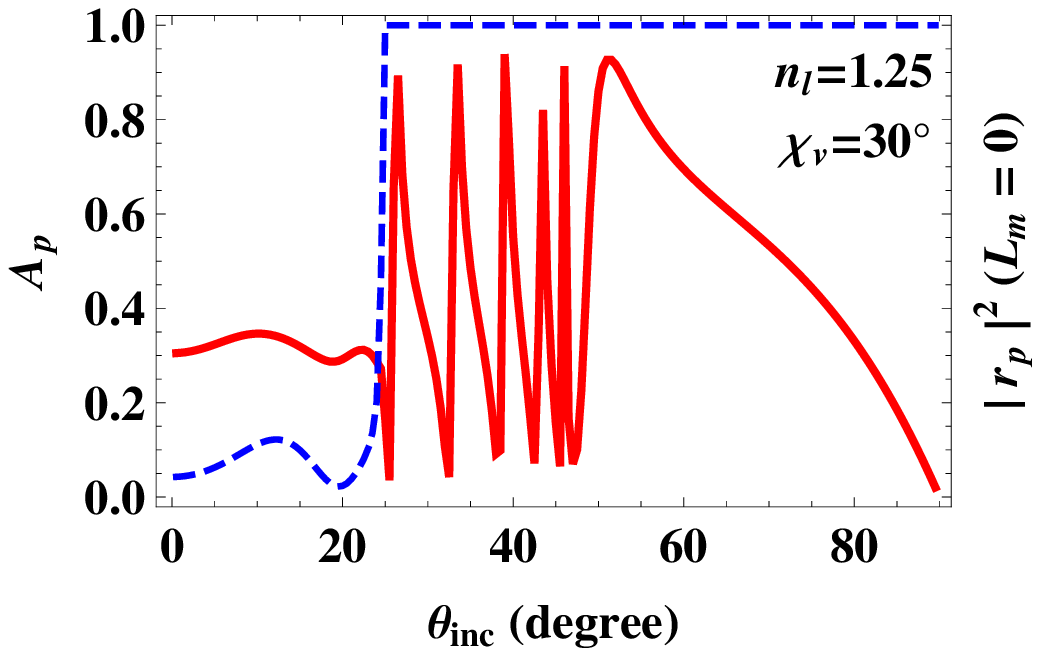}
\includegraphics[width=3.05in]{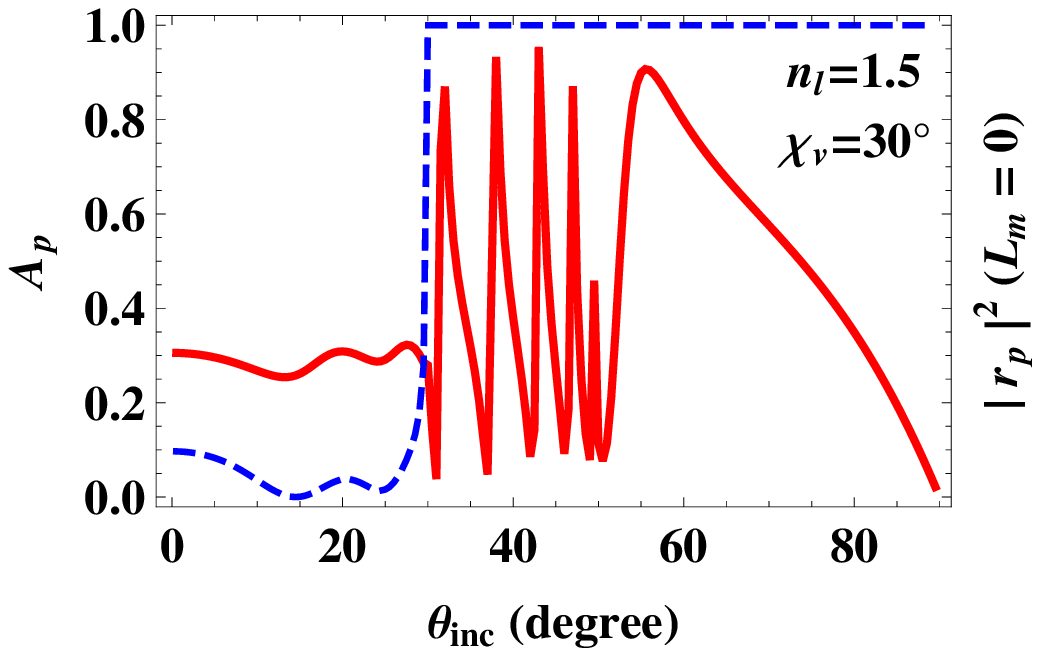}
%\includegraphics[width=2.05in]{ap_rp_100_60.eps}
%\includegraphics[width=2.05in]{ap_rp_125_60.eps}
%\includegraphics[width=2.05in]{ap_rp_150_60.eps}
%\\
%\includegraphics[width=2.05in]{ap_rp_100_90.eps}
%\includegraphics[width=2.05in]{ap_rp_125_90.eps}
%\includegraphics[width=2.05in]{ap_rp_150_90.eps}
 \caption{ \l{Fig3}
 The absorbance $A_p$ (red, solid curve) plotted against $\theta_{inc} $ (in degree),  when $\eps_{d} = 9$, $\eps_m =
 -56 + 21i$, $L_m = 10$ nm, $L_{\Sigma} = L_m + 1000$ nm , $\chi_v = 30^\circ$ and $n_\ell \in \lec 1.0,  1.25, 1.5 \ric$.
  Also plotted is the quantity $|
 r_p |^2$ (blue, dashed curve), calculated when $L_{m} = 0$.
 }
\end{figure}

\newpage

\begin{figure}[!ht]
\centering
\includegraphics[width=3.05in]{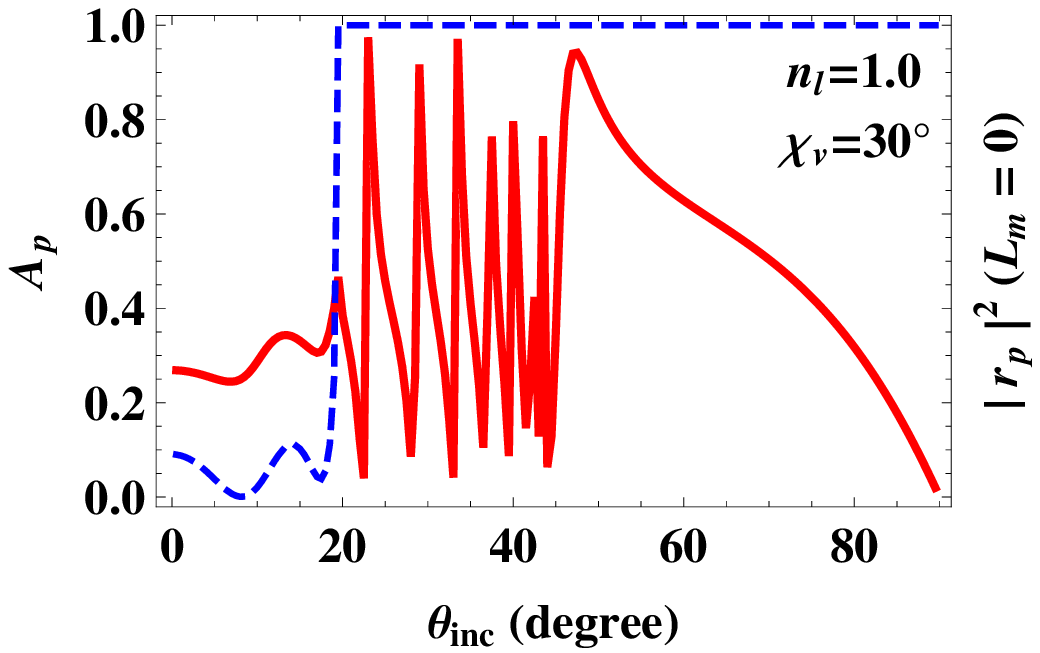}
\includegraphics[width=3.05in]{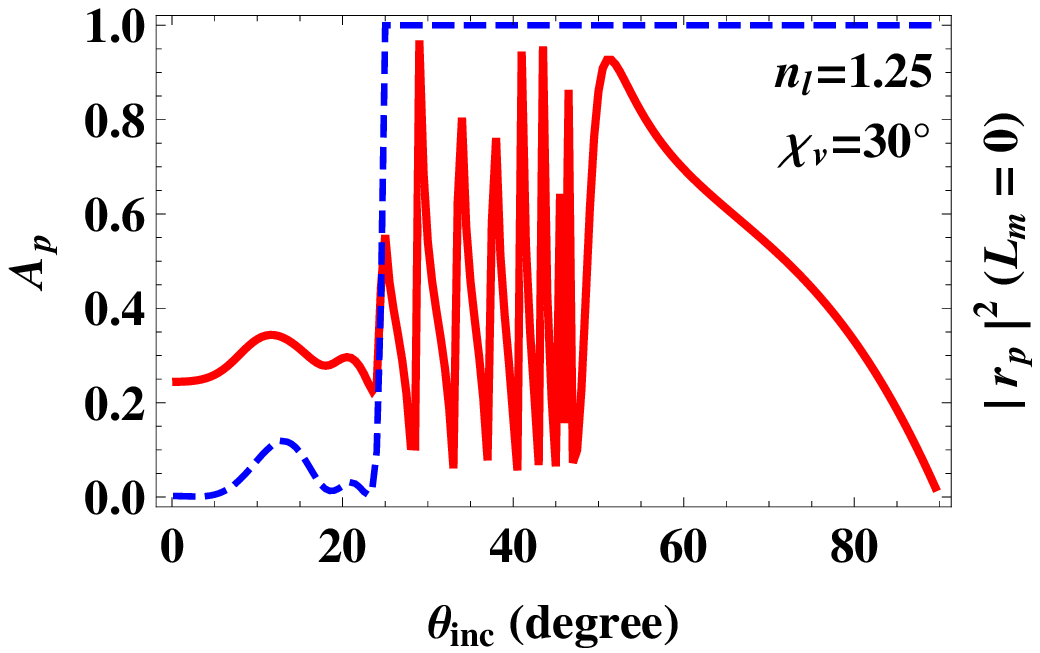}
\includegraphics[width=3.05in]{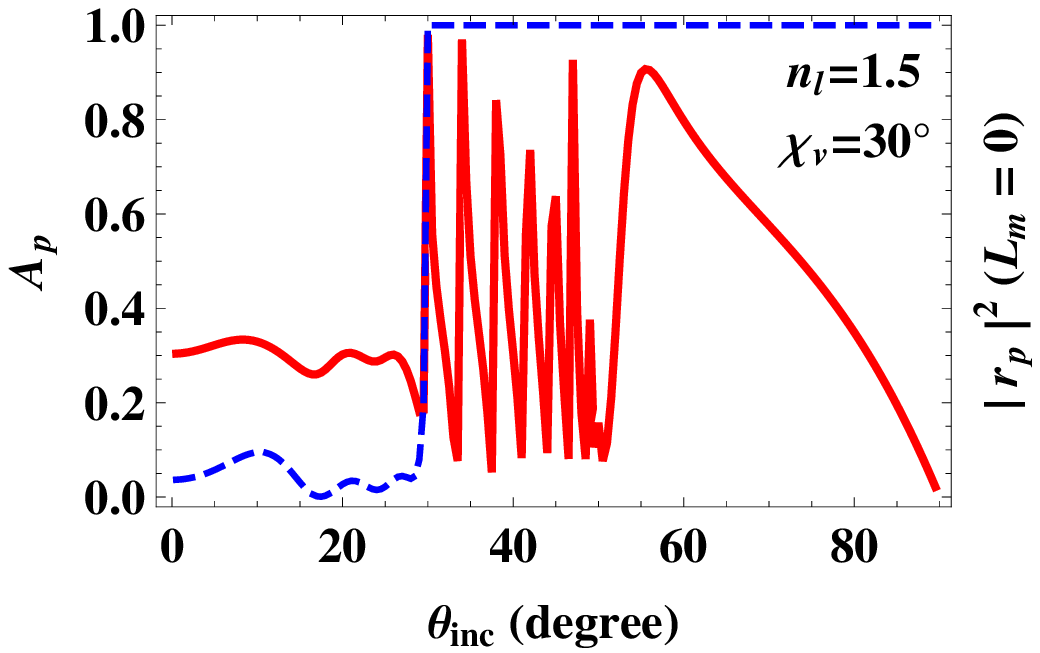}
%\includegraphics[width=2.05in]{L1500_ap_rp_100_60.eps}
%\includegraphics[width=2.8in]{L1500_ap_rp_125_60.eps}
%\includegraphics[width=2.05in]{L1500_ap_rp_150_60.eps}
%\\
%\includegraphics[width=2.05in]{L1500_ap_rp_100_90.eps}
%\includegraphics[width=2.8in]{L1500_ap_rp_125_90.eps}
%\includegraphics[width=2.05in]{L1500_ap_rp_150_90.eps}
 \caption{ \l{Fig3_L1500}
 As Fig.~\ref{Fig3} but with $L_{\Sigma} = L_m + 1500$ nm.
 }
\end{figure}

\newpage

\begin{figure}[!ht]
\centering
\includegraphics[width=3.5in]{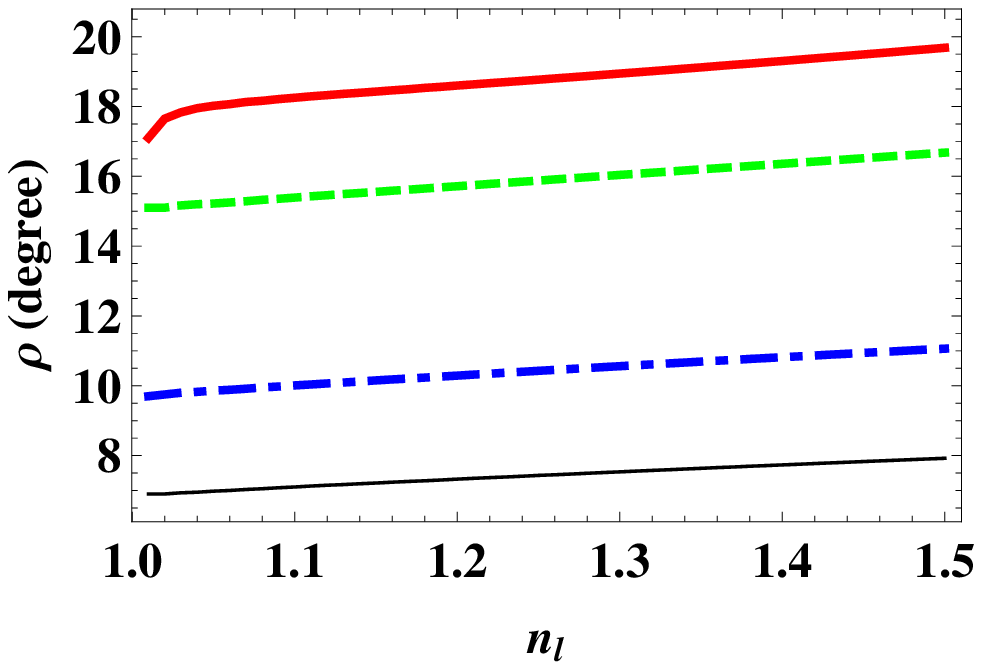}
 \caption{ \l{Fig4}
The figure of merit $\rho$ (in degree) plotted against $n_\ell \in (1.0, 1.5)$
 for $L_{\Sigma} = L_m + 1000$ nm with $\chi_v = 15^\circ$ (red, thick solid curve), $30^\circ$ (green, dashed curve), $60^\circ$ (blue, broken dashed
 curve) and  $ 90^\circ$ (black, thin solid curve).}
\end{figure}


\vspace{10mm}

\begin{thebibliography}{99}

\bibitem{Green2000}
R.J. Green, R.A. Frazier, K.M. Shakesheff, M.C. Davies,
C.J. Roberts, S.J.B. Tendler, Surface plasmon resonance analysis of dynamic biological
interactions with biomaterials, Biomater.  21 (2000) 1823--1835.

\bibitem{Homola2003}
J. Homola,
Present and future of surface plasmon resonance biosensors,
Anal. Bioanal. Chem. 377 (2003) 528--539.

\bibitem{Abdulhalim2008a}
I. Abdulhalim, M. Zourob, A. Lakhtakia,
Surface plasmon resonance for biosensing: A mini--review,
Electromagnetics 28 (2008) 214--242.

\bibitem{Arya2006}
S.K. Arya, A. Chaubey, B.D. Malhotra,
Fundamentals and applications of biosensors,
Proc. Ind. Nat. Acad. Sci. 72 (2006) 249--266.

\bibitem{Luong2008}
J.H.T. Luong, K.B. Male, J.D. Glennon,
Biosensor technology: Technology push versus market pull,
Biotechnol. Adv. 26 (2008) 492--500.

\bibitem{Did}
D. Felbacq, Plasmons go quantum, J. Nanophoton. 2 (2008) 020302.

\bibitem{Kret}
 E. Kretschmann, H. Raether, Radiative decay of nonradiative surface plasmons excited by light, Z. Naturforsch. A 23 (1968) 2135--2136.

\bibitem{Yang1991}
F. Yang, G.W. Bradberry, J.R. Sambles, The study of the optical properties of obliquely evaporated nickel films using IR surface plasmons, Thin Solid Films 196 (1991) 35--46.

\bibitem{Maaroof2007}
A.I. Maaroof, A. Gentle, G.B. Smith, M.B. Cortie, Bulk and surface plasmons in highly nanoporous gold films,
J. Phys. D: Appl. Phys. 40 (2007) 5675--5682.

\bibitem{Abdulhalim2008b}
I. Abdulhalim, A. Lakhtakia, A. Lahav, F. Zhang, J. Xu, Porosity
effect on surface plasmon resonance from metallic sculptured thin
films, Proc. SPIE 7041 (2008)  70410C.

\bibitem{Shalabaney}
A. Shalabney, A. Lakhtakia, I. Abdulhalim, A. Lahav, C. Patzig, I.
Hazek, A. Karabchevsky, B. Rauschenbach, F. Zhang,  J. Xu, Surface
plasmon resonance from metallic columnar thin films, Photon.
Nanostruct. Funda. Appli. (doi:10.1016/j.photonics.2009.03.003).

\bibitem{Sprokel1}
G.J. Sprokel, The reflectivity of a liquid crystal cell in a surface plasmon experiment,
Mol. Cryst. Liq. Cryst. 68 (1981) 39--45.

\bibitem{Sprokel2}
G.J. Sprokel, R. Santo, J.D. Swalen, Determination of the surface tilt
angle by attenuated total reflection, Mol. Cryst. Liq. Cryst. 68 (1981) 29--38.

\bibitem{Lloyd1988}
J.P. Lloyd, C. Pearson, M.C. Petty, Surface plasmon resonance studies of gas effects
in phthalocyanine Langmuir--Blodgett films, Thin Solid Films 160 (1988) 431--443.

\bibitem{Kano}H. Kano, W. Knoll, Locally excited surface-plasmon-polaritons for thickness measurement
of LBK films, Opt. Commun. 153 (1998) 235--239.

\bibitem{Baba}
A. Baba, F. Kaneko, K. Shinbo, K. Kato, S. Kobayashi, Evaluation of
liquid crystal molecules on polyimide LB films using attenuated
total reflection measurement, Thin Solid Films 327-329 (1998)
353--356.


\bibitem{HW}
I.J. Hodgkinson, Q.H. Wu, Birefringent Thin Films and
Polarizing Elements,  World Scientific, Singapore, 1998.

\bibitem{STF_Book}
A. Lakhtakia, R. Messier, Sculptured Thin Films:
Nanoengineered Morphology and Optics, SPIE Press, Bellingham, WA,
USA, 2005.


\bibitem{LP_AJP}
A. Lakhtakia, J.A. Polo Jr., Morphological influence on
surface--wave propagation at the planar interface of a metal film
and a columnar thin film, Asian J. Phys. 17 (2008) 185--191. (The value of $\mbox{Im} \, \les \varkappa \ris$ for $\chi_v = 5^\circ$ presented in
Table~1  should be $0.0108$.)

\bibitem{PL_OC}
J.A. Polo Jr., A. Lakhtakia,
Morphological effects on surface-plasmon-polariton waves at the planar interface
of a metal and a columnar thin film, Opt. Commun. 281 (2008) 5453--5457.




\bibitem{Barreca}
D. Barreca, A. Gasparotto, C. Maccato, C. Maragno, E. Tondello, E. Comini,
G. Sberveglieri, Columnar CeO${}_2$ nanostructures for sensor application,
Nanotechnology 18 (2007) 125502.



\bibitem{PH07}
S.M. Pursel,  M.W. Horn, Prospects for nanowire
sculptured--thin--film devices,  J. Vac. Sci. Technol. B
 25 (2007) 2611--2615.


 \bibitem{ML_inverse_homog}
T.G. Mackay, A. Lakhtakia, Determination of constitutive and
morphological parameters of columnar thin films by inverse
homogenization,  $\mathsf{http://arxiv.org/abs/0909.5375}$



\bibitem{HWH_AO}
I. Hodgkinson, Q.h. Wu,  J. Hazel, Empirical equations for the
principal refractive indices and column angle of obliquely deposited
films of tantalum oxide, titanium oxide, and zirconium oxide,
Appl. Opt.  37 (1998)  2653--2659.

\bibitem{Lakh_Opt}
A. Lakhtakia,
 Enhancement of optical activity of chiral sculptured thin films
by suitable infiltration of void regions,
 Optik 112 (2001) 145--148; corrections: 112 (2001) 544.


\bibitem{MTR1976}
R. Messier, T. Takamori, R. Roy, Structure-composition
variation in rf--sputtered films of Ge caused by process parameter
changes, J. Vac. Sci. Technol.  13 (1976) 1060--1065.



\bibitem{BMYVM}
J.R. Blanco, P.J. McMarr, J.E. Yehoda, K. Vedam, R. Messier,
Density of amorphous germanium films by spectroscopic
ellipsometry, J. Vac. Sci. Technol. A  4 (1986) 577--582.

\bibitem{WRL03}
F. Walbel, E. Ritter,  R. Linsbod, Properties of
TiO${}_{\mbox{x}}$ films prepared by electron--beam evaporation of
titanium and titanium suboxides, Appl. Opt.  42 (2003)
4590--4593.







\bibitem{BET1938}
S. Brunauer, P.H. Emmett, E. Teller, Adsorption of gases in
multimolecular layers, J. Am. Chem. Soc.  60 (1938) 309--319.

\bibitem{BHBP}
G. Bomchil, R. Herino, K. Barla, J.C. Pfister, Pore size
distribution in porous silicon studied by adsorption isotherms,
J. Electrochem. Soc.  130 (1983) 1611--1614.

\bibitem{RHLP}
J.V. Ryan, M. Horn, A. Lakhtakia,  C.G. Pantano,
Characterization of sculptured thin films, Proc. SPIE
5593 (2004) 643--649.


\bibitem{WLM}
W.S. Weiglhofer, A. Lakhtakia,  B. Michel, Maxwell Garnett and
Bruggeman formalisms for a particulate composite with bianisotropic
host medium,  Microwave Opt. Technol. Lett. 15 (1997) 263--266;
corrections:    22 (1999) 221.

\bibitem{EAB}
T.G. Mackay, A. Lakhtakia, Electromagnetic Anisotropy and Bianisotropy: A Field Guide, Word Scientific, Singapore, 2009.



\bibitem{Tait}
R.N. Tait, T. Smy, M.J. Brett, Modelling and characterization of columnar
growth in evaporated films, Thin Solid Films 226 (1993) 196--201.

\bibitem{Welford1987}
K.R. Welford, J.R. Sambles, M.G. Clark,
Guided modes and surface plasmon-polaritons
observed with a nematic liquid crystal using attenuated total reflection,
Liq. Cryst. 2 (1987) 91--105.




\end{thebibliography}
\end{document}